\documentstyle[12pt]{article}
\newcommand{\draft}{
\newcount\timecount
\newcount\hours \newcount\minutes  \newcount\temp \newcount\pmhours

\hours = \time
\divide\hours by 60
\temp = \hours
\multiply\temp by 60
\minutes = \time
\advance\minutes by -\temp
\def\hour{\the\hours}
\def\minute{\ifnum\minutes<10 0\the\minutes
            \else\the\minutes\fi}
\def\clock{
\ifnum\hours=0 12:\minute\ AM
\else\ifnum\hours<12 \hour:\minute\ AM
      \else\ifnum\hours=12 12:\minute\ PM
            \else\ifnum\hours>12
                 \pmhours=\hours
                 \advance\pmhours by -12
                 \the\pmhours:\minute\ PM
                 \fi
            \fi
      \fi
\fi
}
\def\fullclock{\hour:\minute}
\begin{centering}
{\Large \tt Draft: \today, \clock}
\end{centering}
\vskip -1.7cm
$\phantom{a}$
} 
\catcode`\@=11 
\def\lsim{\mathrel{\mathpalette\@versim<}}
\def\gsim{\mathrel{\mathpalette\@versim>}}
\def\@versim#1#2{\vcenter{\offinterlineskip
        \ialign{$\m@th#1\hfil##\hfil$\crcr#2\crcr\sim\crcr } }}
\catcode`\@=12 

\def\myunderline#1{{\em #1}}

\def\etal{{\em et~al.}}
\def\MSbar{$\overline{\hbox{MS}}$}

\def\beq{\begin{equation}}
\def\eeq{\end{equation}}
\def\bea{\begin{eqnarray}}
\def\eea{\end{eqnarray}}
\def\bq{\begin{quote}}
\def\eq{\end{quote}}

\def\NP{{\it Nucl.Phys.} }
\def\PL{{\it Phys.Lett.} }
\def\PR{{\it Phys.Rev.} }
\def\PRL{{\it Phys.Rev.Lett.} }

\def\gappeq{\mathrel{\rlap {\raise.5ex\hbox{$>$}}
{\lower.5ex\hbox{$\sim$}}}}

\def\lappeq{\mathrel{\rlap{\raise.5ex\hbox{$<$}}
{\lower.5ex\hbox{$\sim$}}}}

\begin{document}

\begin{titlepage}
\begin{flushright}
CERN-TH/95-155\\
TAUP-2284-95\\
OSU RN309\\
hep-ph/9509312
\end{flushright}
\bigskip

\begin{centering}
{\large {\bf Pad\'e Approximants, Borel Transforms and
Renormalons: the Bjorken Sum Rule as a Case Study}} \\
\vspace{.3in}
{\bf John Ellis}\\
\vspace{.05in}
Theory Division, CERN, CH-1211, Geneva 23, Switzerland. \\
e-mail: johne@cernvm.cern.ch \\
\vspace{0.5cm}
and \\
\vspace{0.5cm}
{\bf Einan Gardi and Marek Karliner}\\
\vspace{.05in}
School of Physics and Astronomy
\\ Raymond and Beverly Sackler Faculty of Exact Sciences
\\ Tel-Aviv University, 69978 Tel-Aviv, Israel
\\ e-mail: gardi@post.tau.ac.il, marek@vm.tau.ac.il
\\
\vspace{0.5cm}
and\\
\vspace{0.5cm}
{\bf Mark A. Samuel}\\
\vspace{.05in}
Department of Physics, Oklahoma State University, \\
Stillwater, Oklahoma 74078, USA\\
e-mail: physmas@mvs.ucc.okstate.edu   \\ 
\vspace{0.5cm}
{\bf Abstract} \\
\vspace{.05in}
\end{centering}
\vspace{0.5cm}
{\small
We prove that Pad\'e approximants yield increasingly accurate
predictions of higher-order coefficients in QCD perturbation
series whose high-order behaviour is governed by a
renormalon. We also prove that this
convergence is accelerated if the perturbative series is Borel
transformed. We apply Pad\'e approximants and Borel
transforms to the known perturbative coefficients for
the Bjorken sum rule. The Pad\'e approximants reduce considerably
the renormalization-scale dependence of the perturbative correction
to the Bjorken sum rule.
We argue that the known perturbative series is already
dominated by an infra-red renormalon, whose residue we extract and
compare with QCD sum-rule estimates of higher-twist effects.
We use the experimental data on the Bjorken sum rule to extract
$\alpha_s(M_Z^2) = 0.116_{-0.006}^{+0.004}$,
including theoretical errors due to the finite order of available
perturbative QCD calculations,
renormalization-scale dependence and higher-twist effects.
} 
\vfil
\begin{flushleft}
CERN-TH/95-155\\
September 1995 \\
\end{flushleft}
\end{titlepage}
\newpage
\baselineskip=18pt

\section{Introduction}

Everybody interested in more precise quantitative tests of QCD, or in
its
place in some Grand Unified Theory, would welcome a more precise
determination of the strong coupling strength $\alpha_s$ in some
well-defined
renormalization prescription, say \MSbar, at some reference
energy
scale, say $M_Z$.  Such determinations are normally made using
perturbative
QCD to interpret data, though lattice QCD may also become competitive
once systematic effects are better controlled.
Obtaining the desired level of precision using perturbative QCD requires
calculations beyond the next-to-leading order.  Several processes are
calculated to high order in perturbative
QCD \cite{Bjcorr},\cite{HighOrder}.
 However, progress in the
high-precision determination of $\alpha_s(M_Z)$ is hampered by the fact
that
the QCD perturbation series is expected to be asymptotic:
\beq
S(x) = \sum^\infty_{n=0} c_n x^n~, \quad x \equiv \frac{\alpha_s}{\pi}~,
c_n
\simeq n!K^n {n}^\gamma
\label{one}
\eeq
for some coefficients $K, \gamma$ \cite{renmvz}.
Under these circumstances, how can
one
extract the most information from the QCD perturbation series, and
obtain the
best value of $\alpha_s = \pi x$?  The usual answer is to calculate up
to
order $n_{opt}:  \Delta_n \equiv |c_n x^n|$ is minimized, and use the
magnitude $\Delta_{n_{opt}}$ of this minimum term as an estimator for
the
residual uncertainty.

In this paper we study whether
it is possible  to estimate $S_{opt}(x) \equiv
\mathop{\sum}\limits^{n_{opt}}_{n=0}c_n x^n$
reliably without the labour of calculating
all
the perturbative coefficients $c_n:  n \leq n_{opt}$,
and comment on the possibility of summing
over the higher-order terms
$S_{asy}(x)
\equiv \sum\limits^\infty_{n_{opt}} c_n x^n $
sufficiently reliably to reduce the
magnitude of the residual uncertainty below $\Delta_{n_{opt}}$.
Our approach to these issues is based on
Pad\'e approximants, which are widely appreciated in other areas of
physics,
and which we have recently shown \cite{SEK}
can be used to predict higher-order
perturbative QCD coefficients in agreement with exact calculations
(where available) and with the effective charge
method \cite{EffectiveCharge},\cite{KataevStarshenko}.

In this paper,
we present new results on the rate of convergence of Pad\'e
approximants for series of the form (\ref{one}) expected in QCD.
We also
demonstrate that they reduce significantly the renormalization-scale
dependence of the perturbative series for the Bjorken sum rule, and
summarize a comparison with another technique for treating
higher-order effects in perturbative QCD \cite{BLM}.
To go further, we transform to the Borel plane, where behaviours of the
type
(\ref{one}) correspond to discrete renormalon
singularities \cite{renmvz}. The Pad\'e
technique is \myunderline{a priori} well adapted to locating such
singularities, and we indeed prove that the convergence of the Pad\'e
approximants is accelerated for the Borel transform of a series such as
(\ref{one}).  We apply this combined Pad\'e--Borel technique to the
calculated
QCD perturbation series for the Bjorken sum rule, and show that it
yields a leading infra-red renormalon pole
close to the expected location in the Borel plane.
Assuming this location, we extract its pole residue and use it to
evaluate the possible magnitude of the
infra-red
renormalon ambiguity in the perturbative contribution to
the Bjorken sum rule, which we argue is canceled by a corresponding
ambiguity in the non-perturbative contribution.
We use Pad\'e summation,  extracting
$\alpha_s(M_Z^2) = 0.116_{-0.006}^{+0.004}$
from the available polarized
structure function measurements, including theoretical errors
associated with renormalization-scale dependence and higher-twist
effects. The accuracy of this result testifies to the utility of both
Pad\'e Approximants and the polarized structure function data.

\section{``Convergence" of Pad\'e Approximants}

We denote {\em Pad\'e Approximants} (PA's)
to a generic perturbative QCD series $S(x) =
\mathop{\sum}\limits^\infty_{n=0} c_n x^n$ by
\beq
[N/M] = \frac{a_0 + a_1x + ... +a_Nx^N}{1 + b_1x + ... + b_Mx^M}~:~
[N/M] = S + O(x^{N+M+1})
\label{two}
\eeq
i.e. the PA's are constructed so that their Taylor expansion
up to and including order $N{+}M$ is identical to the original series.
We have previously pointed out that the {\em next} term in the Taylor
expansion of a $[N/M]$ PA typically provides increasingly accurate
estimate $c^{est}_{N+M+1}$ of the next
higher-order perturbative coefficient \ $c_{N+M+1}$ of the
original series. In the following we refer to such estimates as
 {\em Pad\'e Approximant Predictions} (PAP's).

Let us briefly restate the condition \cite{PAPconvergence}
for the convergence of PAP's.
With \ $f(n) \equiv \ln c_n$\
and \ $g(n) \equiv
d\,^2{\kern-0.2em}f(n)/dn^2$ \
(where the derivative with respect to $n$ is to be
understood in a discrete sense),
a sufficient condition for PAP convergence,
\ $c^{est}_{N+M+1} \,\rightarrow c_{N+M+1}$, \
is
\ $\lim\limits_{n \rightarrow \infty} g(n) = 0$.
\ To quantify the rate of convergence, we introduce the quantity
\beq
\epsilon_n \equiv \frac{c_{n}\, c_{n+2}}{c^2_{n+1}} - 1 = e^{g(n)} - 1
\label{three}
\eeq
It is easy to check that $\epsilon_n \simeq 1/n$ for an asymptotic
series of the form (\ref{one}).
When the asymptotic behaviour of $\epsilon_n$ is known,
it is possible to write down an asymptotic formula for the
relative error $\delta_{[N/M]}$ in the PAP
estimate $c^{est}_{N+M+1}$ of the
perturbative coefficient $c_{N+M+1}$.
\beq
\delta_{[N/M]} \equiv \frac{c^{est.}_{N+M+1} -
c_{N+M+1}}{c_{N+M+1}}
\label{four}
\eeq
When $\epsilon_n \simeq 1/n$, as is the case
for an asymptotic series of the form (\ref{one}), we have been able to
demonstrate, for all values of $N$ and
several values of $M$, that
\beq
\delta_{[N/M]} \simeq
-\,\frac{M!}{L^M  }\,, \quad~{\rm where}~ L = N+M+a'M+b~.
\label{five}
\eeq
for some choice of numbers $a', b$. This implies that
asymptotically
\beq
\ln \big|\,\delta_{[M/M]} \,\big|\simeq - M \,[1 + \ln (2 + a')]
\label{fiveone}
\eeq
We have verified that this formula is numerically accurate for simple
series of the form (\ref{one})
in which $c_n = n! K^{n} n^\gamma$. Moreover, we have checked the
prediction
(\ref{fiveone}) for several different asymptotic series, including that
for the QCD
vacuum polarization $D$ function in the large-$N_f$ approximation
\cite{Dfunction},
where it agrees numerically very well with
the
relative error reported in panel (a) of the figure in Ref. \cite{SEK}.
We note that the large-$N_f$
$D$ function contains an infinite number of renormalon poles, which
could in general provide important corrections to the leading-order
formulae
\hbox{(\ref{five}, \ref{fiveone})}. The fact that this is not
the case supports the empirical utility of the PAP's even
beyond the idealized analytical case (\ref{one}).

In the previous paragraph, we have discussed the use of Pad\'e
approximants to estimate
the next term in a given perturbative series, and, as we have seen,
sufficient
conditions for the convergence of such PAP's are known, so that the main
open issue is
their actual rate of convergence in practical applications. Next we
discuss how to use PA's to estimate the ``sum" of the perturbation
series,
which we term {\em Pad\'e Summation} (PS).

It is important to note that
convergence of the PAP's  is   largely  independent
of the ``summability" of a given series.
Thus, for example, the PAP method gives equally precise predictions
for the next terms in the series
\ $\sum\limits_0^\infty n! \, x^n$
\ and
\ $\sum\limits_0^\infty n! \, (-x)^n$\ ,
even though the latter is Borel summable, while the former is not.

The formal issues related to Pad\'e Summation are less clear,
especially because
most perturbation series of practical interest are not Borel summable
(since their Borel
transforms have poles on the positive real axis).
One well-defined prescription for defining the ``sum" of such a series
is the Cauchy
principal value of the inverse Borel transform
integral \cite{PvaluePrescription}, so we ask
whether PS
``converges" to this prescription for the ``sum" of the series. Such
series are in
general obtainable from functions with cuts on the positive real axis,
with a toy
example being provided by the simple asymptotic series \
$\sum\limits_0^\infty n! \, x^n$,
which is a formal expansion of
\beq
\int^\infty_0~~{e^{-t}\over 1-xt}~~dt = {1\over x}
\int^\infty_0~~{e^{-y/x}\over
1-y}~~dy
\label{seven}
\eeq
We see from the representation on the right-hand side of (\ref{seven})
that this
series corresponds in
QCD language to a single simple infrared
renormalon pole, whilst
the left-hand side of (\ref{seven}) exhibits a cut on the positive
real axis. Figure 1 exhibits the errors with respect to the
Cauchy principal value of the integral (\ref{seven})
of conventional partial sums
and PA's to the series $ \Sigma_0^{\infty} n! x^n$ for the case
$x=0.1$ for which $n_{opt}=9$.
We see that the first few PA's have relative errors that are
considerably smaller than those of the partial sums.
The best PA has an error comparable to $\Delta n_{opt}$, despite
using as input a number of terms that is less than $n_{opt}$.
However, we see that higher-order PA's exhibit relative errors
that are irregular, and may even be less accurate than the
conventional partial sums. This is because the
Pad\'e method mimics the cut on the left-hand side of (\ref{seven})
with an ever-denser set of poles, the $[N/M]$ PA's having typical
separations $\Delta x = {\cal O}(1/M)$. This makes it ever more
difficult to avoid nearby
poles when the PA's are evaluated at any fixed value of $x$, and
therefore to define
``convergence" of the PS procedure. However, PA's of sufficiently high
order are again much more accurate than the partial sums, which are
blowing up because of the renormalon singularity.

As also seen in Figure 1, it is possible to improve on the simple PA's.
One way of smoothing out the irregularities associated with nearby poles
is to evaluate
the real parts of the $[N/M]$ PA's off the real
axis\footnote{We thank A.A. Migdal for pointing this out.}
at $x + i\epsilon$.
It is a theorem \cite{BenderOrszag}
that this smoothed PS prescription
converges to the
Cauchy principal value
 in the cut plane $|\,\hbox{arg}\ z\,| > 0$,
if the weight in the integral over the positive
real axis is
positive. Another approach is simply to remove from the PA any
spurious contribution of a
nearby pole. Yet another is to make a Taylor expansion of the PA
and truncate it at order $n_{opt}$. We have developed \cite{next}
criteria for deciding which of these is better for any given case,
and the application to our toy example is also shown in Figure 1,
designated by ``combined method".
Since the investigation of these techniques
is still in progress, we do not discuss them further in this paper,
deferring this to a future publication
\cite{next},
where a detailed description will
be given, together with physical applications.

\section{Application to the Bjorken Sum Rule}

We now discuss a concrete application of PA's
to a perturbative QCD series, namely that for the Bjorken sum rule,
which takes the following form in the \MSbar\
renormalization prescription \cite{BjI},\cite{Bjcorr}:
\beq
\int^1_0 [ \,g_1^p(x,Q^2) - g_1^n(x,Q^2)\,] =
{1 \over 6}\, |g_A| \, f(x)\,\,:
\quad
f(x) = 1 - x - 3.58 x^2 - 20.22 x^3 + ... + (HT)
\label{bjf}
\eeq
for $N_f = 3$, as relevant to the $Q^2$ range of current
experiments, where $x = \alpha_s(Q^2)/\pi$, the dots
represent uncalculated higher orders of perturbation theory,
and $(HT)$ denotes higher-twist terms. The perturbative series in
(\ref{bjf}) is expected to be dominated by renormalons in large
orders \cite{largeNfBjSR},
leading to growth in the perturbative coefficients $c_n^{Bj}$
of the form shown in (\ref{one}). The PA's to the series
(\ref{bjf}) yield the following predictions for the next term
\bea
c^{Bj}_{4[PA]}\approx {-}111\qquad (\,\hbox{[1/2] \ PA)}
\nonumber\\
\label{bjpa}\\
c^{Bj}_{4[PA]}\approx {-}114\qquad (\,\hbox{[2/1] \ PA)}
\nonumber
\eea
in this series.

In order for these PAP's to be useful, it is
important to estimate the errors involved.
The asymptotic error estimate given in eq.~(\ref{five})
requires as input values of $a'$ and $b$, which are not known
a priori. Experience with many series shows that typically
${-1} \lsim a' \lsim 0$ and $b\approx 0$.
With $a'=0$ and $b=0$ we obtain a ballpark estimate
\beq
\delta_{[1/2]} \simeq {-}2/9; \qquad
\delta_{[2/1]} \simeq {-}1/3
\label{bje}
\eeq
In a previous paper \cite{SEK},
we used a different method to estimate the errors of the Pad\'e
prediction (\ref{bjpa}), obtaining $c^{Bj}_{4[PA]}={-112}\,\pm\,33$
as the error-weighted average of the [1/2] and [2/1] approximants.
We also pointed out  that this prediction
is close to an estimate made using the Effective Charge
Method (ECH):
\beq
c^{Bj}_{4[ECH]} \simeq {-}130
\label{bjech}
\eeq
We now note further that even though the error estimates (\ref{bje})
are in principle expected to hold
only asymptotically, in practice
(\ref{bjpa}) and (\ref{bjech}) are consistent with each other within
these estimates.
We take this as an indication that the true value of $c_4^{Bj}$
is likely to be in the range predicted by the PAP and ECH methods.
Moreover, as we shall show in section 4, it seems that the
perturbation series in (\ref{bjf}) is already dominated by a
single infrared renormalon, in which case the quantities
(\ref{bje}) are
no longer ``statistical" errors, but fractional corrections to be
subtracted from (\ref{bjpa}), improving the
concordance with the ECH estimate (\ref{bjech}).

The next step is to apply the PS procedure to estimate the complete
correction
function $f(x)$ in (\ref{bjf}). Figure 2 compares $f^{[2/2]}(x)$, the
[2/2] PS estimate of $f(x)$
(obtained using the ECH value (\ref{bjech}) of the fourth-order
perturbative coefficient),
\beq
f^{[2/2]}(x)=
{\frac { 1- 8.805\,x+ 11.974\,{x}^{2}}{ 1- 7.805\,x+
 7.753\,{x}^{2}}}
\label{f22}
\eeq
with the
[1/2], [2/1] PS's
and with the partial sums of
the perturbative series up to order $x^3$ and $x^4$ (the latter also
taken from (\ref{bjech})\ ).
 We see that the different
 PS's are numerically quite stable
in the range $x
\lappeq 0.1$ of relevance to present experiments, which is related to
the fact that the
nearest poles are some distance away
($x = 0.18$ for the $[2/1]$ PA, $x=-3.41$ and $x = 0.18$
for the
$[1/2]$ PA,
and $x=0.15$ and $x=0.86$ for the [2/2] PA).
This means that the ``combined method" for
smoothing of PA's described at the end of the
previous section is not necessary.

We now check the reliability of the PS estimates of $f(x)$ in two
different ways, first
by checking their renormalization-scale dependences.
 Figure 3 compares the values of $\alpha_s(Q^2)$ estimated from a
fixed value of $f(x)$ at $Q^2 = 3$ GeV$^2$,
when one renormalizes at scales
$\mu$ in the range between $Q/2$ and $2Q$, using the PS or the
third-  or fourth-order partial sum.  We see that the value of
$\alpha_s(Q^2)$ extracted from the PS is indeed
much less
$\mu$-dependent than the values extracted using the
partial sums \cite{ScaleDependence},
consistent with our belief that
     the PS's provides a reliable
estimate  of the full correction factor $f(x)$,
which should be independent of $\mu$. In particular,
the $\mu$-dependence of the [2/2] PS
is small compared with other possible sources of
theoretical error.

We have also compared the PS's with results based on the BLM
treatment of the perturbative series in the \MSbar\ scheme, in which
the growing higher-order coefficients are absorbed into the
scales $Q^2$ at which $x=\alpha_s(Q^2)/\pi$ is evaluated in each of
the lower-order terms.
As described elsewhere \cite{next},
we find that our PS procedure agrees
very well with the BLM procedure when applied to the perturbative
series for the Bjorken sum rule, adding further support to our
contention
that the PA's may indeed accelerate usefully the convergence of
perturbative QCD series, as suggested by the general arguments of
Section 2.

\section{Pad\'e Approximants in the Borel Plane}

If one knows the asymptotic behaviour of the series under study, one can
go
further.  In particular, if the perturbative coefficients diverge as in
(\ref{one}), which is believed to be the case in perturbative QCD,
corresponding to a discrete set of renormalons \cite{renmvz},
it is useful to consider
the Borel transform
of the series $S(x)$ in eq.~(\ref{one}):
\beq
\tilde S(y) \equiv \sum^\infty_{n=0} \tilde c_n y^n~:~\tilde c_n =
\frac{c_{n+1}}{n!}\,\left({4\over \beta_0}\right)^{n+1}\,\,;
\qquad \beta_0 = (33- 2\,N_f)/3
\label{six}
\eeq
If the Borel transform $\tilde S(y)$ indeed has a discrete set of
renormalon
singularities $r_k/(y-y_k)^P$,
where the $r_k$'s are the residues,
 PA's in the Borel plane are {\em a priori} well
suited to
find them.  Indeed, if there is a finite set of renormalon
singularities, as
occurs for the Bjorken sum rule series in the large-$N_f$ approximation,
higher-order PA's will be {\em exact}.
   In general, the removal of the $n!$
factors
in the coefficients (\ref{six}) means that the corresponding
quantity measuring the rate of convergence of the Pad\'e prediction
for the next term
is
\beq
\tilde \epsilon_n = \frac{\tilde c_n \tilde c_{n+2}}{\tilde c^2_{n+1}}
\simeq \frac{1}{n^2}~,
\label{sevenone}
\eeq
which is much smaller than the previous
$\epsilon_n  \simeq 1/n$ (\ref{three}).  This means that
the
relative error $\tilde \delta_{[N,M]}$ in the PA of the
Borel-transformed
series is also much smaller than (\ref{five}) for the original series,
\beq
\tilde \delta_{[M/M]} \simeq - \,\frac{(M!)^2}{L^{2M}}
\label{eight}
\eeq
corresponding asymptotically to
\beq
\ln\big|\,\tilde \delta_{[M/M]}\,\big| \simeq - 2M [1 + \ln(2 + a')]
\label{eightone}
\eeq
We have also checked this prediction for several different asymptotic
series,
including that for the QCD vacuum polarization $D$ function in the
large-$N_f$ approximation \cite{Dfunction},
which has a discrete infinity of renormalon
poles.  The prediction (\ref{eightone}) again agrees numerically
very well with
the relative error reported in panel (b) of the figure in
Ref.~\cite{SEK},
which
is much smaller than that for the naive PA in panel (a). Again, the
success of the prediction (\ref{eight}) gains significance
from the fact that the large-$N_f$ calculation exhibits an
infinity of renormalon poles, indicating that the Borel PA's are
useful in the real world, and not only in idealized simplified
situations.

We now apply this combined Borel/Pad\'e technique to the QCD
perturbation series (\ref{bjf}) for the Bjorken sum rule.
With the normalization of the Borel variable $y$ implicitly defined
through eq.~(\ref{six}),
the [2/1] PA to the Borel transform of (\ref{bjf}) has a pole at
$y$ = $1.05$ with residue \ $r$ = $0.98$, as seen in Fig. 4.
The appearance of a pole near $y = 1$ is
encouragingly consistent with the exact large-$N_f$
calculations \cite{largeNfBjSR},
which yield
poles at $y = \pm 1, \pm 2$.  In the \MSbar\ prescription
that we are
using, the residues of these poles contain factors exp$[5y/3]$.
Therefore, it
is not surprising that an infra-red renormalon pole at $y = 1$ emerges
more
clearly than an ultraviolet renormalon pole at $y = - 1$.
{\it Prima facie}, the
message of this analysis is that the calculated Bjorken series is
already
dominated by the expected leading infra-red renormalon.

 Encouraged by this success,
we have made fits to the Borel transform of the Bjorken series with
varying
numbers of poles whose locations are fixed in accordance with
theoretical expectations.  The residues found in these
various fits are also
plotted in Fig. 4, where we see the following points:
\begin{itemize}
\item[(i)] the residue of the $y = 1$ pole is consistently found to be
{\it positive} and around unity,
\item[(ii)] the residue of the $y = - 1$ pole is {\it much smaller},
and consistent with zero,
\item[(iii)] there is room for a second pole at $y = 2$,
but it is not possible to
disentangle this from a higher-lying pole.
\end{itemize}

We now discuss the possible implications of this Borel/Pad\'e
exercise for phenomenology. First, we note that the dominance by
a single infrared renormalon pole indicates that, as already
remarked, the fractional errors (\ref{bje}) should be {\it subtracted}
from the naive estimates (\ref{bjpa}), bringing them into better
agreement with the ECH estimate (\ref{bjech}):
\beq
c_{4[PA]}^{Bj} \simeq -136, -152
\label{padecorr}
\eeq
from the [1/2] and [2/1] PA's respectively.
Secondly, it is well known that the magnitude
of the residue $r_1$, of the $y = 1$ pole corresponds to a possible
renormalon
ambiguity $\pm \pi r_1$ relative to the Cauchy principal value discussed
previously. This is also shown for our toy series in Fig.~1.
Taking $r_1$ for the Bjorken series
from Fig.~4, we find an ambiguity \beq
\Delta\left(\Gamma^p_1 - \Gamma^n_1\right) = \pm \frac{|g_A|}{6}\,
0.98\,\pi\,\frac{\Lambda^2}{Q^2}
\label{bjrenamb}
\eeq
in the perturbative contribution to the Bjorken sum rule.
Numerically,
with $\Lambda = 250\pm50$ MeV,
this corresponds to $\pm 0.040\pm0.016 \ {\rm GeV^2}/Q^2$, which
is to be compared with previous QCD sum rule higher-twist estimates that
yield \cite{HTrefs}
\beq
\Delta_{HT}\,\left(\Gamma_1^p - \Gamma_1^n\right)
= - \,\frac{0.02\pm0.01}{Q^2}
\label{bjht}
\eeq
(see also the discussion pertaining to eq.~(7) in Ref.~\cite{BjSRalphas},
and a recent estimate in Ref.~\cite{BraunMoriond} ).
We note that the order of magnitude of the renormalon
ambiguity we find (\ref{bjrenamb})
is close to the higher-twist calculation (\ref{bjht}).

Since the full QCD prediction for any physical quantity must be unique,
the renormalon ambiguity
must be cancelled by a corresponding ambiguity in the definition of the
higher-twist term.
For this reason we do not interpret the renormalon
ambiguity as leading directly to an ambiguity in $\alpha_s$,
but rather use the uncertainty in the higher twist term
(\ref{bjht}).

\section{Extraction of $\alpha_s$ from Bjorken Sum Rule Data}

We conclude this paper by extracting \cite{BjSRalphas}
$\alpha_s(M_Z^2)$ from data on the Bjorken sum rule,
including a discussion of
theoretical errors. We combine the available experimental
evaluations of
$\Gamma_1^{p,n}(Q^2)$ to obtain
\beq
\Gamma_1^p(3 {\rm GeV}^2) - \Gamma_1^n(3 {\rm GeV}^2) = 0.164 \pm 0.011
\label{N}
\eeq
where we have evolved  all the experimental results
\cite{oldSLACa}-\cite{Roblin}
 to the common reference scale $Q^2 = 3$ GeV$^2$.
Evolution of the proton data from 10
to 3 GeV$^2$ is based on the analysis in
Ref.~\cite{Grenier}. The corresponding evolution of the
deuteron data is based on the proton data, combined
with the expected $Q^2$ dependence of the Bjorken sum rule, with
$\alpha_s(Q^2)$ needed as input
determined via iteration, in a self-consistent way.
Evolution to 3 GeV$^2$
from values of $Q^2$ different from 10 GeV$^2$
was treated in a linear approximation to the full
dependence of the data on $1/\log(Q^2)$, which is adequate
for this purpose.

The range of $f(x)$ corresponding to (\ref{N}) is shown as a
vertical error bar in Figure 2.
We use the
[2/2] PS (\ref{f22}) to obtain:
$\alpha_s (3~{\rm GeV}^2) = 0.328^{{+}0.026}_{{-}0.037}$.
We evolve this up to $M_Z^2$ by numerical integration of
the three-loop $\beta$ function \cite{HighOrder}, locating
the $b$-quark threshold in the \MSbar\ scheme at $m_b=4.3\pm0.2$ GeV
\cite{RPP}, and using the three-loop matching conditions of
ref.~\cite{Bernreuther}, to find
\beq
 \alpha_s(M_Z^2) = 0.119_{{-}0.005}^{{+}0.003}\,\, \pm \,\,\dots~\qquad,
\label{N+2}
\eeq
where the $\pm \,\,\dots$ \ in (\ref{N+2})
recalls that theoretical errors remain to be
assigned.

There is a
theoretical error associated with the spread in the different
evaluation procedures shown in Fig. 2,
which we estimate from the difference between the [2/2] and
[1/2], [2/1] PS's to be
$\Delta_{proc}\alpha_s(3~{\rm GeV}^2) = \pm 0.014$,
corresponding to
\beq
\Delta_{proc}\alpha_s(M_Z^2) = \pm 0.002
\label{N+3Z}
\eeq
Another way of looking at the theoretical error uses the
$\mu$-dependence
of the [2/2] PS
shown in Fig. 3 to
estimate
$\Delta_{\mu}\alpha_s(3~{\rm GeV}^2) = \pm  0.009$,
corresponding to
\beq
\Delta_{\mu}\alpha_s(M_Z^2) = \pm 0.001
\label{N+4Z}
\eeq
from varying $\mu$ between $Q/2$ and $2Q$. Both (\ref{N+3Z}) and
(\ref{N+4Z}) are estimates of the uncertainty in $\alpha_s(M_Z^2)$
due to our
uncertainty in the functional form of the QCD correction factor $f(x)$,
so that it could be
regarded as double counting to include them both. Nevertheless, to be
conservative we will add
them in quadrature.\footnote{We have also considered possible systematic
theoretical errors due to
uncertainties in the evolution of $\alpha_s$ up to $M_Z$, including
unknown higher-order terms in the QCD $\beta$-function, the uncertainty
in $m_b$, and freedom in treating the heavy-flavour
threshold \cite{Bernreuther}. They
contribute an error in $\alpha_s(M_Z,2)$ which is much
less than $\pm 0.001\,$.}

We also include a shift and error in the
determination of $\alpha_s(Q^2)$
inferred from the estimated range (\ref{bjht}) of the higher-twist
correction:
$\Delta_{HT}\alpha_s (3~{\rm GeV}^2) = {-}0.024 \pm 0.014$,
corresponding to
\beq
\Delta_{HT}\alpha_s(M_Z^2) = - 0.003 \pm 0.002
\label{N+5Z}
\eeq
Combining (\ref{N+2}), (\ref{N+3Z}), (\ref{N+4Z}), (\ref{N+5Z}),
we extract
\beq
\alpha_s (M_Z^2) = 0.116_{-0.005}^{+0.003} \pm 0.003 ~,
\label{N+6}
\eeq
where the first errors are the experimental errors in (\ref{N+2}), and
the second errors are the
sums in quadrature of the errors in (\ref{N+3Z}), (\ref{N+4Z}),
(\ref{N+5Z}).

Our final value of $\alpha_s(M_Z^2)$ (\ref{N+6}) is
compatible with the central value extracted from compilations
of previous measurements \cite{qcdrev},
and has an error which is competitive.
As well as the experimental error,
we have included motivated estimates of a number of theoretical errors,
using information obtained from our study of Pad\'e approximants. We
believe that this exercise demonstrates the utility of Pad\'e
Approximants in QCD, and the value of polarized structure function
data for determining $\alpha_s(M_Z^2)$.

\bigskip

{\bf Acknowledgments}

We thank
David Atwood,
Bill Bardeen,
Stan Brodsky,
Georges Grunberg,
Sasha Migdal
and Al Mueller
for useful discussions.
This research was supported in part by the Israel Science Foundation
administered by the Israel Academy of Sciences and Humanities,
and by
a Grant from the G.I.F., the
German-Israeli Foundation for Scientific Research and
Development.
It was also in part supported by the US
Department of Energy under Grant No. DE-FG02-94-ER40852.

 \newpage
 \medskip
\noindent
 {\bf FIGURE CAPTIONS}\\

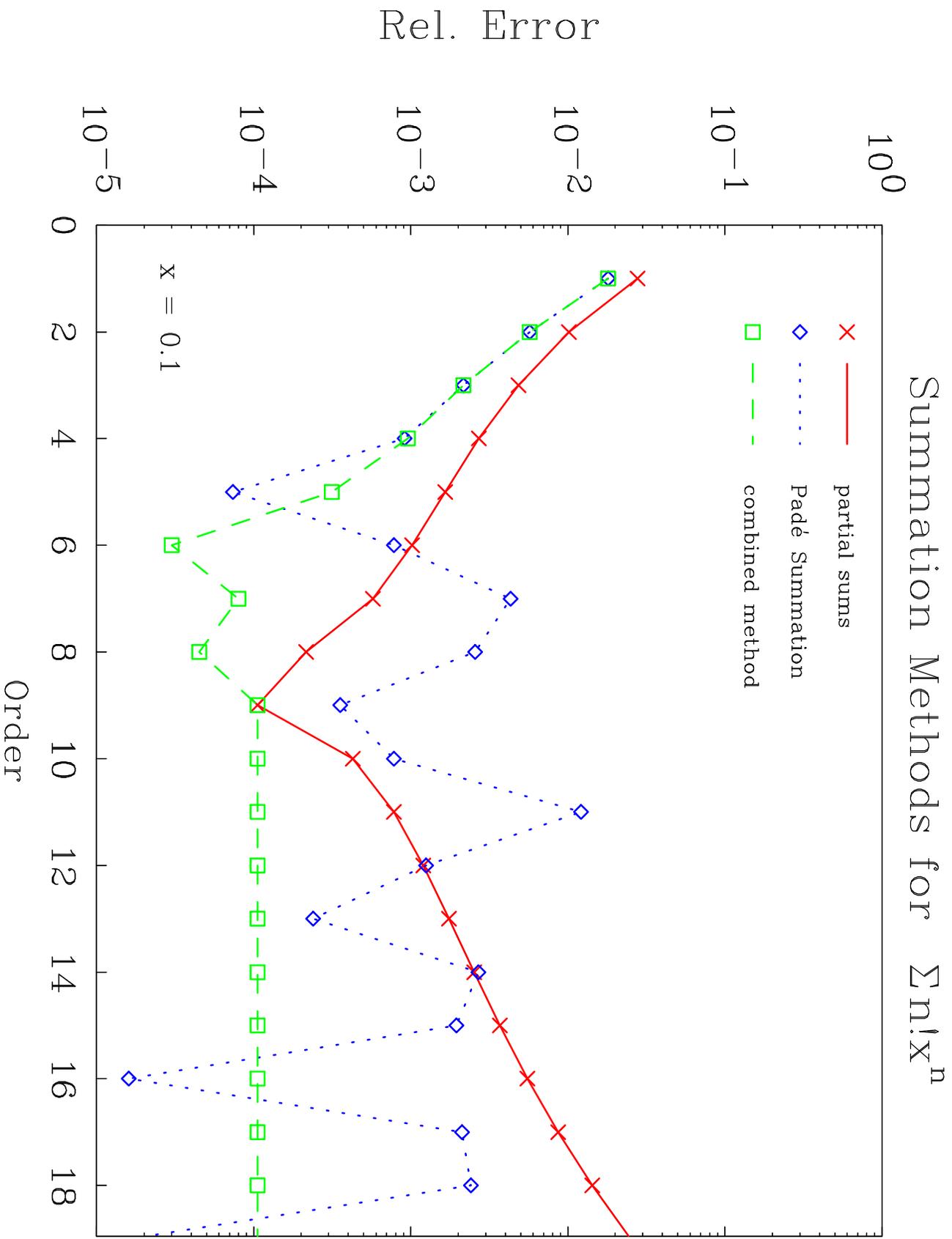
{\bf Fig. 1.}
The relative errors between partial sums of the series
$S(x) = \Sigma n! x^n$ and the Cauchy principal value of the series
(solid line)
is compared with the relative errors of Pad\'e Sums (dotted line).
We see that the relative errors of the Pad\'e Sums are smaller
than those of the partial sums in
low orders, fluctuate in an intermediate r\'egime, and are again
more accurate than the partial sums in higher orders. The
fluctuations are associated with nearby poles in the Pad\'e Sums,
that may be treated by the ``combined method"
mentioned in the text, shown as the dashed line.

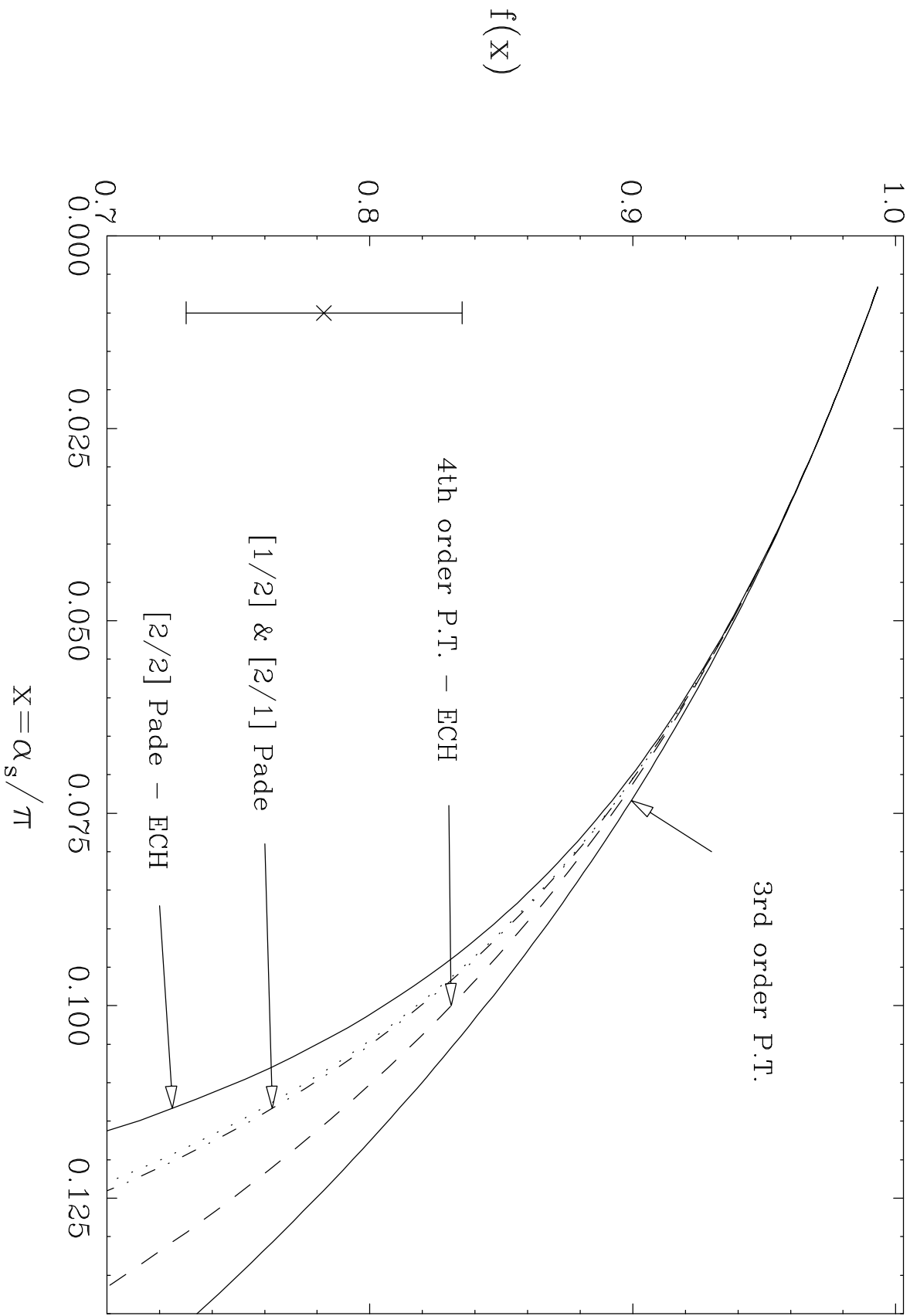
{\bf Fig. 2.}
Different approximations to the Bjorken sum rule correction
factor $f(x)$, third-order and fourth-order perturbation theory,
[1/2], [2/1] and [2/2] Pad\'e Sums (\ref{f22}),
are compared. Also shown
as a vertical error bar is the value of $f(x)$
we extract from the available polarized structure data
(\ref{N}).

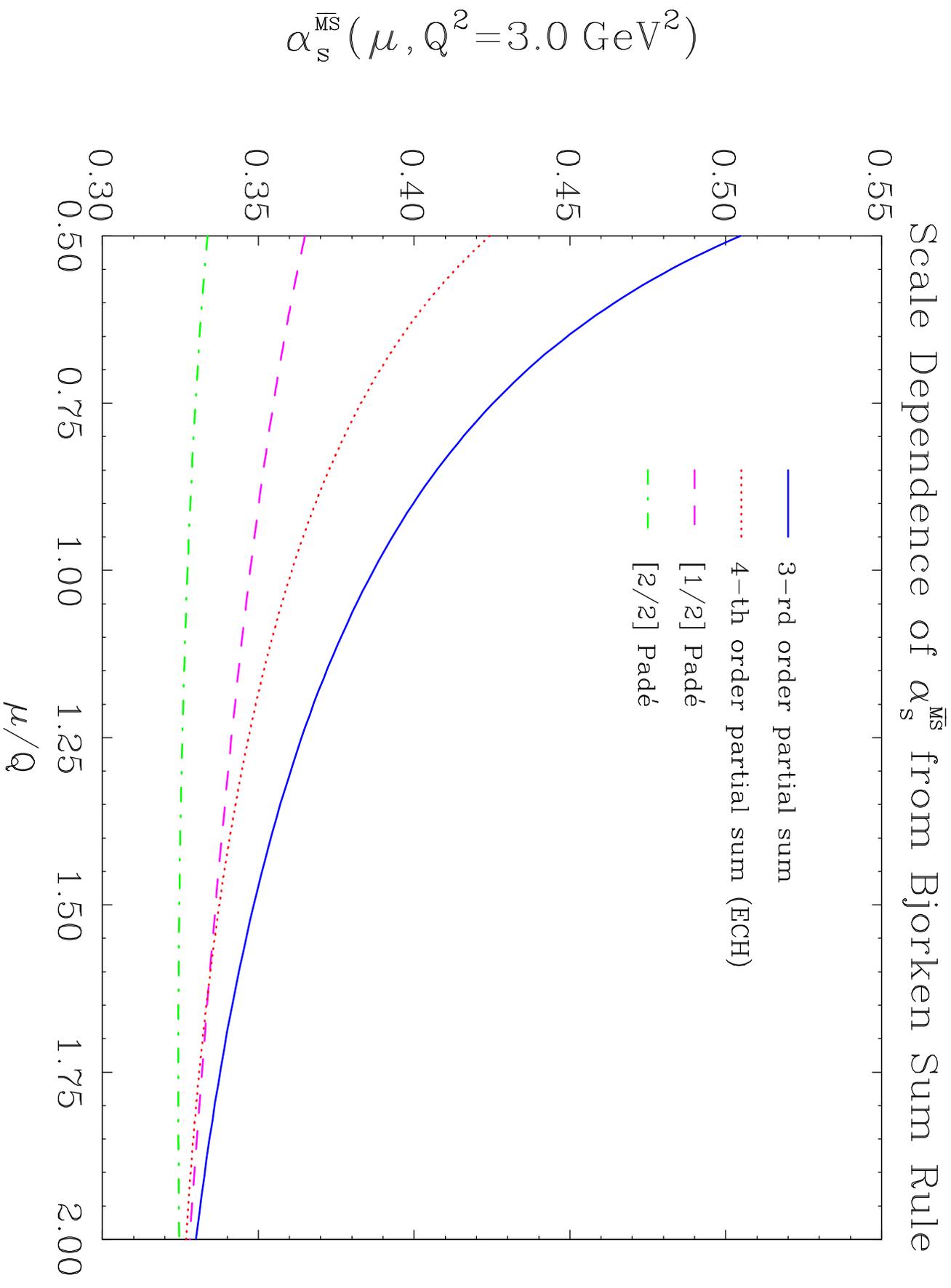
{\bf Fig. 3.}
The scale dependence of $\alpha_s(3 {\rm GeV}^2)$ obtained from
a fixed value
$f(x) = $ $(6/ g_A) \times 0.164 =0.783$,
({\em cf.} eq.~(\ref{N})\ ),
for $Q/2 < \mu < 2Q$, using the naive third- and fourth-order
perturbative series and the [1/2], [2/1] and [2/2] PS's.

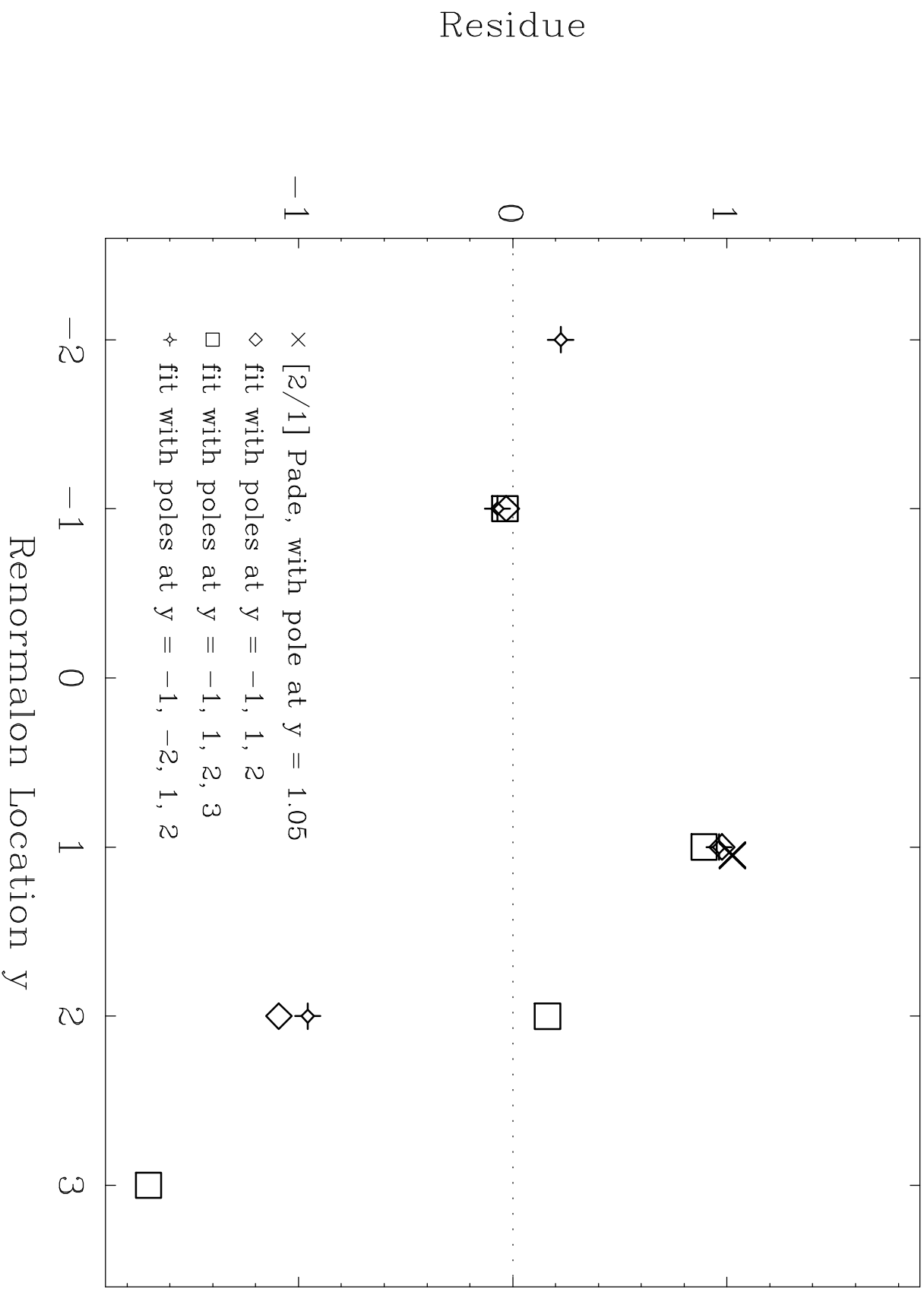
{\bf Fig. 4.}
The locations and residues of
poles in the [2/1] PA and in rational-function
fits to the Borel transform of
the first four terms in the perturbation series for the Bjorken sum rule.
We note that the location of the lowest-lying infrared renormalon
pole is estimated accurately by Pad\'e Approximants in the Borel
plane, and that its residue is stable in the different fits.

\end{document}